# Detecting Straggler MapReduce Tasks in Big Data Processing Infrastructure by Neural Network

Amir Javadpour[1], Guojun Wang[1,*], Samira Rezaei[2] and Kuan-Ching Li[3]

[1]School of Computer Science, Guangzhou University, Guangzhou, China, 510006
[2] Bernoulli Institute for Mathematics and Computer Science, University of Groningen, Netherlands
[3]Department of Computer Science and Information Engineering, Providence University, Taiwan
*Correspondence to: csgjwang@gzhu.edu.cn

*Abstract*— Straggler task detection is one of the main challenges in applying MapReduce for parallelizing and distributing large-scale data processing. It is defined as detecting running tasks on weak nodes. Considering two stages in the Map phase (copy, combine) and three stages of Reduce (shuffle, sort and reduce), the total execution time is the total sum of the execution time of these five stages. Estimating the correct execution time in each stage that results in correct total execution time is the primary purpose of this paper. The proposed method is based on the application of a backpropagation Neural Network (NN) on the Hadoop for the detection of straggler tasks, to estimate the remaining execution time of tasks that is very important in straggler task detection. Results achieved have been compared with popular algorithms in this domain such as LATE, ESAMR and the real remaining time for WordCount and Sort benchmarks, and shown able to detect straggler tasks and estimate execution time accurately. Besides, it supports to accelerate task execution time.

## I. INTRODUCTION

Information technology is becoming a vital part of human life[1]–[4]. Increasing data production and the rapid growth of information technology over the past two decades have led to the production of a considerable amount of data in various formats from various sources such as RFID tag[5], weblogs, data on scientific researches such as healthcare [6], network management[7]–[12]. Several studies have been conducted to model data and partition the query loads on several hosts[13]. In [14], the authors focused on collecting information in wireless sensor networks in cloud structure within a limited time. Authors in [15] propose a new kind of deep neural network which produces interpretable representation in the hidden layer.

Hadoop is an open-source software framework that is released in 2003, and works similar to an operating system processing and managing large volumes of data on different machines. Hadoop core consists of a Hadoop distributed file system storage and a MapReduce processing [16]. This article is based on Yet Another Resource Negotiator (YARN) framework introduced by Yahoo in 2010 [17]. As in Figure 1 and adapted from [18], it summarizes different research areas in Hadoop. This research is focused on task scheduling. Among all the methods for task scheduling, we focused on dynamic adaptive schedulers based on speculative execution. Some of the new methods in the literature related to this category are ESAMR [20] and SECDT [25]. Gray boxes in figure 1 represent the focused area of this article.

MapReduce process as Big Data processing infrastructure "chops" jobs into several tasks. For organizations with large data volume, it is challenging to use the traditional database management system to process and analyze data. By design of MapReduce, millions of users worldwide can find their content across the internet within a fraction of second. The strength of this method is due to the parallelization of these two phases (Map and Reduce). In fact, MapReduce is an infrastructure for writing distributed programs that use a massive volume of data. Paralleling applications with a distributed environment require programmers to consider many issues[19]. For example, how a large file is, how to split a huge file into small chunks, among others. The input of each mapping task is a small fraction of the total input data. Mapping is responsible for reading input, applying the mapping function, sorting and merging the inputs. In the previous version of Hadoop, the reduction phase did not start until the completion of the mapping phase while in its next-generation (which is also called YARN), it is not necessary to complete the mapping phase before starting to reduce function. In MapReduce structure, tasks are first sent to the mapping machine which has performed in two stages, copy and combine. The reduce phase has three stages: shuffle, sort and reduce.

The total execution time of all the stages constitutes the total execution [19]–[23], [38], [39]. As data is transferred in the copy and shuffle stages, they have the most significant impact at execution time. There is an assigned weight for each stage, which is the ratio of the stage's execution time to the total execution time [22]–[25], [40]. It is also possible to calculate errors by comparing the estimated weights by a real one [15]. The execution time of the complete job depends on the execution of the slowest task. The running task on the slowest node is called a straggler task, which should be detected and assigned to another node by speculative execution[16]. The LATE [19] considers the same weights for each stage in phases, while it is not the case in the experiments. Considering the heterogeneous environment of Big Data processing, it is not efficient to choose the same weight for each stage. ESAMR[20] calculates the weights of each task with the K-means algorithm [19]. Due to the use of an unmanaged algorithm and lack of appropriate parameters for grouping, this method cannot correctly classify performed executions and cannot correctly calculate the remaining execution time.

This paper proposes a dynamic method for estimating execution time which can also be used in heterogeneous environments. The neural networks are applied to estimate the weights according to task execution time. The purpose of this study is to reduce the error of estimating the remaining time from the execution of tasks and correctly identifying straggler tasks to increase the efficiency in the Big Data computational infrastructure. The remaining of this article is organized as follows. Section II reviews the presented algorithms in the literature, and the proposed solutions are introduced in Section III. In the Sections IV and V, the proposed method and the experiments are described and finally, conclusions and ideas for future work in Section VI.

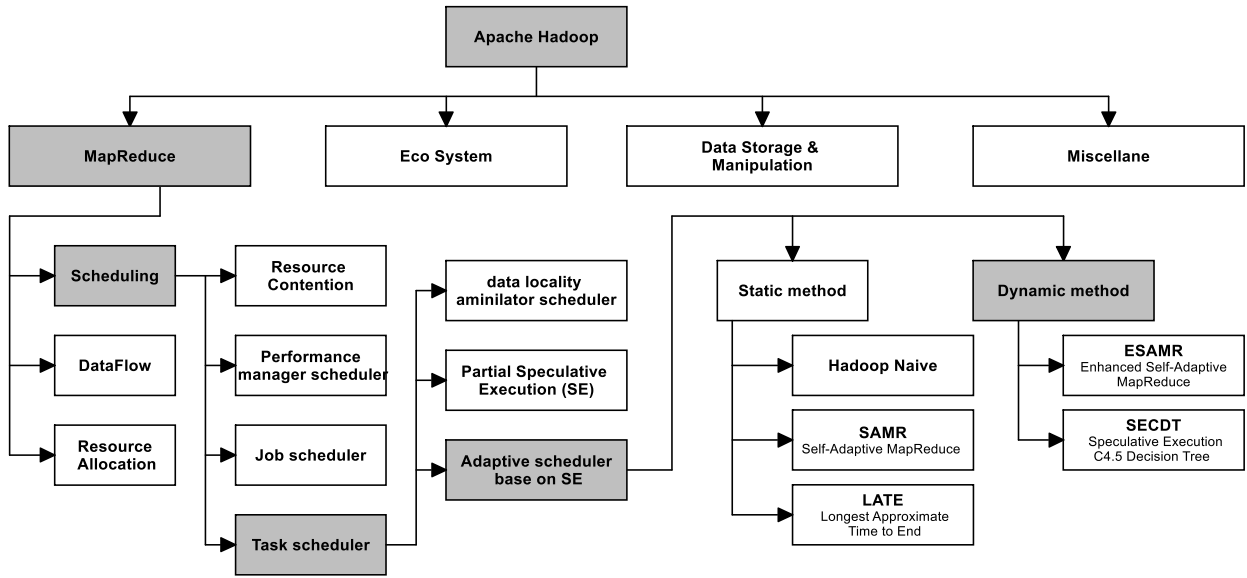

FIGURE 1: FRAMEWORK OF APACHE HADOOP STRUCTURE [18].

## II. LITERATURE REVIEW

The proposed method is based on solving the shortcomings of the current solutions in the literature. For example, in LATE method, the running task remaining time for every phase has been considered the same, while in the Reduce phase, the shuffle stage requires more time to complete compared to other stages. SAMR, ESAMR and SECDT methods are not able to predict the running time correctly as it is based on the previous task. Due to the different characteristics of the previous tasks with the current one, it is not accurate enough. ESAMR merely uses executable information and does not consider node specifications while it is important to take this into account as the node processing times are different based on their characteristics (CPU and memory). In the decision tree-based approach or SECDT, pruning of branches causes information loss and consequently false estimation. In the proposed method based on neural network, with the help of stored running information on previous tasks, finds similar tasks to the currently running one and identifies appropriate weights.

Authors in [30] present a real-time task-scheduling algorithm while considering energy-management requirements. It is based on grouping according to the latest cut-off time, where each group adopts a dynamic optimization strategy to make scheduling decisions. Authors in [13] propose a Comprehensive Transmission (CT) model for reliable data transmission. They also design a Two-Phase Resource Sharing (TPRS) protocol, which mainly consists of a pre-filtering phase and a verification phase, to efficiently and privately achieve authorized resource sharing in the CT model.

### A. HADOOP NAÏVE METHOD

This method uses the Hadoop Naive algorithm to control nodes in Hadoop [28], [31]. In this algorithm, weights in Map and Reduce phase are (M1 = 1, M2 = 0) and (R1=R2=R3=1/3) respectively. The following calculations are performed after the task has for at least one minute. All the related equations in the literature to this article are classified in Table 1. The progress score of a task t is based on how much of a task's (key, value) pairs has been completed. Equations 1 and 2 show the progress score (Ps) of tasks in Map and Reduce phases and it is in range of [0,1]. X is the number of (key, value) pairs that have been processed successfully and Y is the overall number of (key, value) pairs. K is the stage (shuffle, sort, and merge) in a Reduce phase [32]. Equation 3 calculates the average value of $P_s$ and assigns it as the threshold. In this formula, N is the total number of executing tasks. Lastly, equation 4 identifies the straggler task by checking the value of $P_s$ being less than 20% of average $P_s$ [17].

## B. LATE (LONGEST APPROXIMATE TIME TO END) METHOD

The purpose of LATE is to calculate the remaining time of executing tasks [17], [23]. While weights are the same as Hadoop naïve method, Equation 5 selects straggler tasks by considering remaining time. Pr stands for progress rate and t is the consumed time of the task. We use Equation 6 to calculate the remaining time for the execution of tasks in a specific stage (TTE). Tasks are sorted according to their maximum remaining time. 10% of total tasks (known as a speculative cap) are selected as straggler tasks. They will be reassigned to a node which satisfies Equation 7. LATE cannot accurately estimate the remaining time. It uses constant numbers for the weight of each step while the effect of each step for each task differs from others and the ratios are not always constant. In the proposed method, with the help of stored executable information, it finds similar tasks and identifies appropriate weights for each step.

TABLE 1. THE RELATIONSHIPS USED IN THE LITERATURE REVIEW. ALL THE PRESENTED EQUATIONS ARE BASED ON THE FOLLOWING REFERENCES [20], [19], [27], [26]

| FORMULA | REFERENCE | FORMULA | REFERENCE |
|---|---|---|---|
| $P_s = X/Y$ (1) | [32] | slow node threshold= 25 % of all nodes (7) | [28] |
| $P_S = 1/3(K + X/Y)$ (2) | [32] | $APR = \sum_{i=0}^{N} \Pr[i]/N$ (8) | [28][34] |
| $Avg(p_s) = \sum_{i=1}^{N} p_s[i]/N$ (3) | [32] | $\Pr[i] < (1-STaC) \times APR$ (9) | [28] |
| $Ps \leq Avg(p_s) - 20\%$ (4) | [17] | $BackupNum < Bp \times TaskNum$<br>$0 < Bp < 1$ (10) | [28] |
| $pr = Ps/t$ (5) | [28][34] | $ATTE = \sum_{k=1}^{N} \binom{n}{k} TTE[i]/N$ (11) | [28][34] |
| $TTE = (1-Ps)/Pr$ (6) | [28] | $TTE[i] - ATTE > ATTE \times STT$ (12) | [28] |

## C. SAMR (SELF ADAPTIVE MAPREDUCE SCHEDULING ALGORITHM) METHOD

This method stores the stage's weight of performed tasks in XML file and uses it to estimate the next task execution time[26]. The initial stage's weights are defined as (1,0,1/3,1/3,1/3). $P_s$ is calculated every 100 milliseconds. Equation 8 estimates average execution speed and equation 9 finds the straggler task. The close value of STaC to zero demonstrates straggler tasks. The number of backup tasks (BackupNum) is also calculated using equation 10 in which TaskNum is the number of executing tasks. Bp has empirically set to 0.2. Average Time To End of all running tasks (ATTE) is calculated in equation 11. If $i^{th}$ task fulfills equation 12, it is a slow task. Slow Task Threshold (STT) in the range [0,1] is used to measure how fast or slow a task is. In this article, we assume STT =0.4 as STT lower than 0.4 considers many fast tasks to be slow. For STT larger than 0.4, many slow tasks are supposed to be fast.

## D. ESAMR (ENHANCED SELF-ADAPTIVE MAPREDUCE SCHEDULING ALGORITHM)

ESAMR was introduced in 2012 to improve the SAMR algorithm [36]. Besides storing weights of the previous tasks, the K-means algorithm divides this information into k (k=10) categories. After execution, at least 20% of the total tasks in the Map and Reduce stages, a temporary weight will be calculated for that phase. The algorithm searches for a cluster with the closest weight to assign the partly complete tasks to them. If there is no completed task on a node, the average weight of all clusters will be used.

## E. SECDT (SPECULATIVE EXECUTION ALGORITHM BASED ON DECISION TREE)

SECDT [25] considers node characteristics and executive information with the help of the decision tree to find similar nodes, predict the exact execution time faster. CPU speed, free memory, amount of input data and network transmission speed are influential factors in task execution. These characteristics along with weights are stored and used to build the tree iteratively. Tree scrolling will be time-consuming due to the increasing amount of information, so the algorithm clears stored information every three hours. Table 2 compares the proposed method with previous studies. The characteristics of the methods are categorized in the columns of this table. The proposed method in this article is suitable for a heterogeneous environment by saving the weights of each stage for different nodes. The proposed method uses a neural network to estimate the weights of each stage. NN (Neural Network) inputs are weights of other performed tasks. Obtained NN output is used to estimate execution time. In Table 4, the fourth column refers to the use of other features such as the amount of RAM and CPU of each container to estimate execution time. In the proposed method, this item is not considered, since the proposed method uses new executive data to estimate the execution time and by changing the container

`

characteristics the executive information is recorded. Some methods execute all straggler tasks, and such methods are specified in the last column, where the proposed method re-executes 10% of the total number of tasks.

TABLE 2. COMPARISON OF METHODS USED IN CONJUGATION.

| Characteristics and method | Appropriate for heterogeneous environment | Estimation of weights based on previous executed tasks on the system | Estimation of execution time dynamically | Executive specification used | Maximum for number of SE |
|---|---|---|---|---|---|
| Hadoop Naïve [28] | It is not suitable for the heterogeneous environment because it does not use all nodes' information. | It does not use the weights of the executed tasks | The execution time of the running task is compared with the average execution time of the executed tasks. So it is static | The execution time of the executed tasks is considered as input | All tasks that are lower than average execution time are identified as straggler tasks and will execute again |
| LATE [28] | constant coefficients are considered for weights by default and are not suitable for heterogeneous environments. | It does not estimate and use constant weights | Because it uses static coefficients it is not dynamic | The rate of ask progress is used as input to calculate the remaining execution time | Max SE= 10% of total task |
| SAMR [28][34] | It is not suitable for the heterogeneous environment because it uses the previous task coefficients for the new task and does not consider heterogeneity. | It only uses executive information of the previous task | It is not dynamic because it uses coefficients of the previous task for a new task | The previous weights of executed tasks are considered as input for executing new task | Not mentioned in the article. |
| ESAMR [36] | Each node uses the executive information of executed tasks. | The first information of each node is classified then searches about weights | With the help of the k-means algorithm, first weights are estimated then remaining execution time is calculated | Weights of executed tasks are considered as k-means algorithm input | Max SE= 10% of total task. |
| SECDT [36] | Each node uses the executive information of executed tasks. | First, the information of each node is mapped to the decision tree then it searches for weight in each branch | With the help of decision tree algorithm, first weights are estimated then remaining execution time is calculated | Weights of previous tasks and also cpu, and RAM of every container are considered as algorithm input | Not mentioned in the article. |
| Proposed Method | Each node uses the executive information of executed tasks. | The neural network is trained with the received information and then estimates the weights in each node | With the help of an artificial neural network algorithm the weights are first estimated then the remaining execution time is calculated. | Weights of executed tasks and also amount of processed data are considered as Input of Neural Network Algorithm | Max SE= 10% of total task. |

## III. PROPOSED METHOD

The innovations of speculative execution with the ANN framework are divided into two parts. First, the proposed method is capable to be used applied in heterogeneous and homogeneous environments. Second, it estimates the weights of each stage accurately, which results in shorter total execution time. The proposed method uses two categories of variables: dependent (assigned weights to each stage and the remaining execution time of the execution of the MapReduce tasks) and independent (execution time, the amount of processed data and the progress rate) variables. As demonstrated in Figure 2, the proposed method has three main components: Information storage repository for saving information from previously executed tasks, weight estimation by neural network and the estimation of task execution time. Each incoming tasks will be divided into map and reduce phases. The generated reports on this stage are categorized into train and test datasets as input values for NN to estimate weights. The obtained results in test data can also be used in order to learn the model and improve the accuracy of results. By estimating the weights in NN, it is also possible to estimate the remaining time of task execution which is the key to finding straggler tasks.

Figure 3 demonstrates the relationship between the presented components in Figure 2 through a flowchart. By entering a job through the Hadoop, resources will be assigned to this job by the resource manager. The system manager searches for straggler tasks after t seconds. In this step, by using executing information of working nodes, progress score, progress rate and remaining executing time will be calculated for each task in order to find straggler tasks. Tasks are sorted base on the remaining execution time. If the number of speculative tasks is higher than 10 percent of whole tasks, the task with the highest remaining time will be executed.

`

In the proposed method, the user can add a job to the Hadoop which makes the system manager assign resources to the new job. After t seconds of each task running time, the system manager will look for straggler tasks. The next step is to order all the remaining tasks based on their remaining execution time. Slave nodes save the information of speculative running of current jobs. The system manager gets all the information about the currently running tasks and runs NN to this information. Progress rate, remaining time execution and task progression is calculated and if the number of speculative tasks is higher than 10 percent of the total number of tasks. In that case, no changes will be applied to the execution process. Otherwise, the task with higher remaining time will be started. This process will be applied to all new incoming jobs.

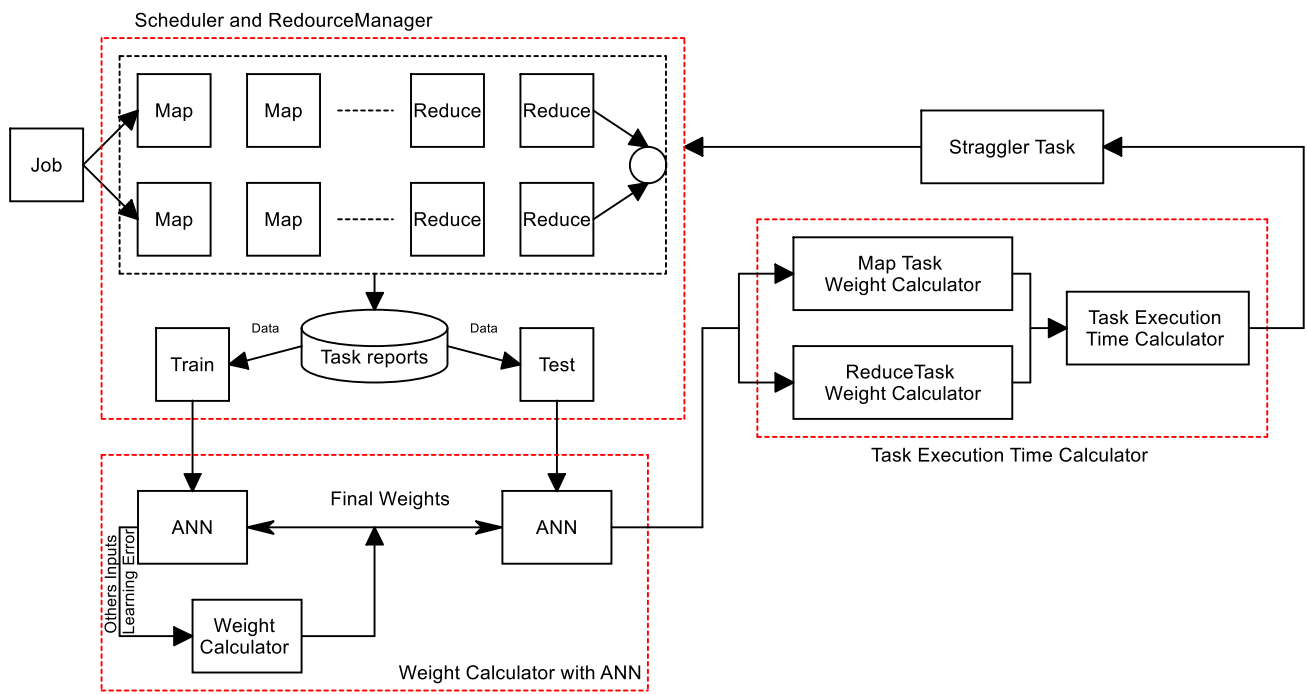

FIGURE 2: PROPOSED FRAMEWORK FOR SPECULATIVE EXECUTION WITH ANN

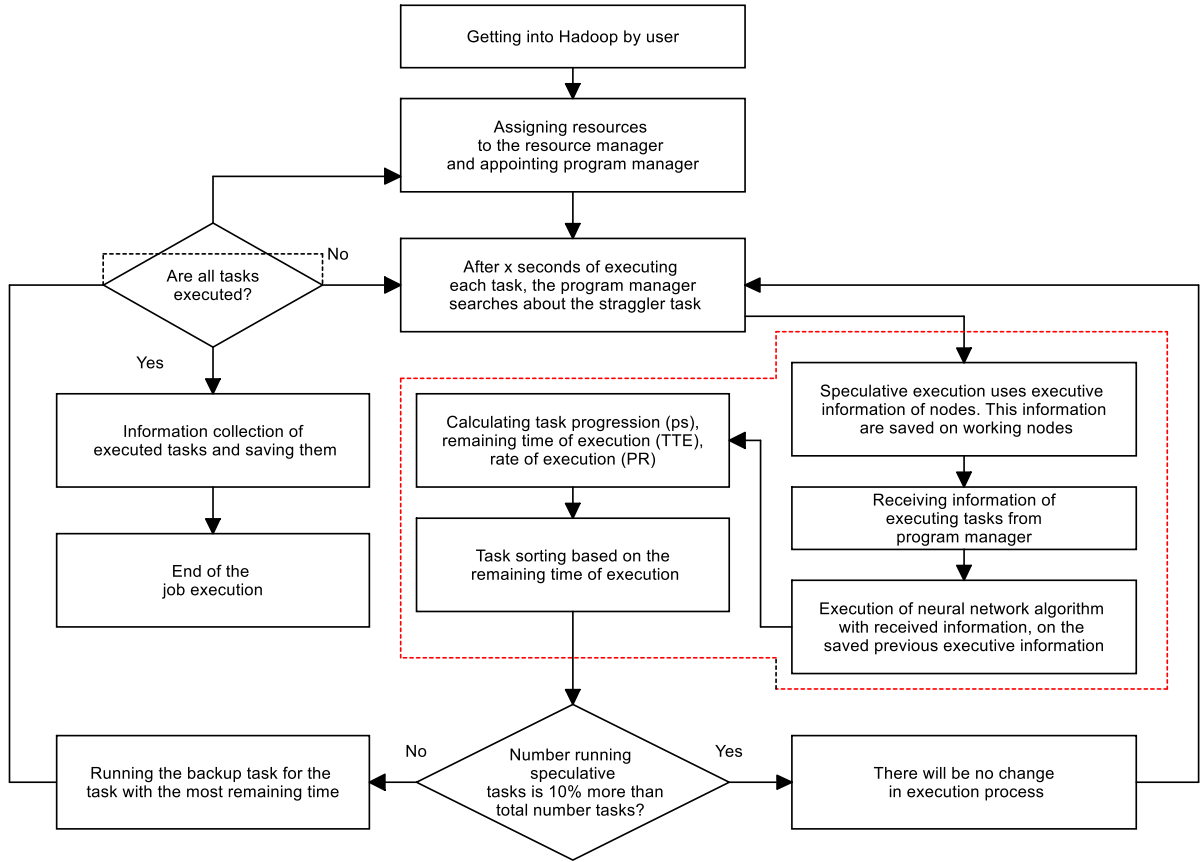

FIGURE 3: FLOWCHART OF PROPOSED METHOD.

We have used the backpropagation neural network in this research to estimate weights. After each task is completed, the ratio of the execution time of each stage (in Map and Reduce) to the total execution time of each phase is stored as the weight of the stage. The neural network uses this stored executable information to estimate runtime. The neural network algorithm implementation process is based on comparing the actual value to the estimated one. Learning in a backpropagation neural network is based on a multilayer feedforward neural network. In each iteration, the training weights change so that the mean square error of the actual value and the estimated value is the least. As it is described in Appendix A, to use the neural network, the first Algorithm A specifies the phase of the currently running task. If the phase is Map, algorithm B and for Reduce algorithm C will be executed. Algorithm A, also calculates the weights of R1, R2, and R3.

- SCHEDULER

The proposed method is designed to improve the speculative execution process. A new incoming job will be broken into MapReduce tasks and each task will be assigned to a node. The task execution report is periodically sent to the source manager and stored in the repository. We use a container on each node to execute assigned tasks. While the tasks are executing, the source manager monitors the execution process to find straggler tasks.

- *Weight calculation using neural network*

In each iteration of the proposed method, the actual weight value is compared with the estimated one. Weights of all three stages of the reduce phase are calculated by the neural network. Backpropagation NN is implemented in this research, in each iteration, the estimated weight will be compared to the real one. Depending on the achieved accuracy, the learning will either continue to the next iteration or will stop with the current estimations. We also have defined the number of epochs as another termination factor. SubPS and $P_s$ will be calculated. $P_s$ will be calculated based on equation 13; k indicates the number of the current stage. Subps is the ratio of processed key/value pairs ($N_f$) to all key/value pairs ($N_a$).

$$P_s = Subps * \sum_{k=1} R_k \qquad (13)$$

$$Subps = \frac{N_f}{N_a} \tag{14}$$

- *Reduce phase*

Estimating the total and remaining execution time is done based on the calculated weights of each stage. The amount of data to be processed and the current task progress level is essential to calculate Subps and execution time. In the speculative execution, the task with the highest execution time is assigned to another node for execution. All straggler tasks do not re-execute and arranged based on the remaining execution time. In the proposed method, the Mapping phase is considered as one stage; hence the neural network with the input of Subps and Nf and remaining execution time is estimated and stored in the TTE.

TABLE 3. PARAMETER INITIALIZATION USED IN THIS ARTICLE

| Characteristic | value |
|---|---|
| Hadoop cluster installation mode | Fully distributed |
| Number of the cluster node | 5 |
| RAM level node1,2 | 4G |
| RAM level node 3,4 | 3 G |
| Network topology | Slave Master |
| The size of virtual machines hard drive | 50 G |
| Master node | Contains a job follower |
| Slave node | Contains a task follower and a data node |
| distributed file system block size | 128 MB |

TABLE 4. EXPERIMENT DESIGN TABLE.

| A: Number of experiment , B: Fixed parameters , C: variable parameters ,D: Explanations | | | |
|---|---|---|---|
| A | B | C | D |
| 1 | Weight of map-reduce phase of executed tasks | Comparing the estimated weight of backup vector machine algorithms and artificial neural network | Learning rate= 0.05 Epoch= 100 |
| 2 | Weight of map-reduce phase of executed tasks | Estimated weight with the help of an artificial neural network algorithm in Map phase | Learning rate= 0.05 Epoch= 100 |
| | | Estimated weight with the help of an artificial neural network algorithm in reduce phase | |
| | | Progress rate of executed tasks | |
| | | Amount of input | 0.251G.13G |
| 3 | Weight of map-reduce phase obtained from the previous experiment | Remaining time of tasks' execution | With the help of the previous experiment, the weights needed for this experiment is obtained |
| 4 | Weight of map-reduce phase of executed tasks | The execution time of the job | Measuring execution time with changing the number of nodes and amount of input |
| | | Number of nodes | 1-5 |
| | | Amount of input Source manager after 60 seconds of execution searches about the straggler task. | 0.251G-13G |

## IV. EVALUATION

The proposed method is implemented on Hadoop 2.7.3, and Ubuntu 16.0 OS on each virtual machine (VM). The simulation machine has 16 G RAM, Intel i7 processor running at 3.2 GHz. Distributed File System HDFS (HDFS) has a web console for monitoring and checking file systems that allow users to view the content of the HDFS file system. All tests are based on three blocks with a block size of the 128 MB HDFS distributed file system. Figure A in the appendix represents how nodes interact with each other. We have considered two slaves for each of the Map and Reduce tasks. Changes have been made to the Hadoop version 2.7.3 speculate section. The results are generated based on the t-word counting program [17], [18], [24] and compared with LATE, No-Speculate and ESAMR based on task execution time, job execution time and stages' weights. WordCount reads text files and counts the number of repetitions for each word. Mapping is responsible for dividing lines into consisting words and generates pairs of key/value for each word. Input dataset is a set of words. We also have compared the result of our presented method with Sort benchmark as it is also widely been used in the literature to compare the quality of proposed methods. The regular block size in Hadoop 2 is 128 MB. When a larger file is inserted into HDFS, it will be braked down into 128 MB pieces and divided between data nodes. Table 3 describes the parameter specifications for the implementation. The implemented cluster is configured with a node manager, a job follower, a master, and four slave nodes. Each slave hosts a data node and task follower. We consider four different experiments to measure errors on weight estimation, task execution time and job execution time. The first experiment contains a comparison between the proposed method and SVR. The second experiment calculates weights for Map and Reduce phases using recorded information in the execution process. The third experiment uses obtained weights to calculate the remaining execution time. The final experiment is designed to measure the effect of speculative execution in the volume of data and the number of different nodes on the execution time. Table 4 demonstrates the constant and variable parameters of experiments in a neural network.

`

*1) First Experiment*

Regarding the continuous essence of weights, regression methods such as support vector machines and artificial neural networks are efficient in estimating them. In the proposed method, the parameters are set based on error minimization while SVR uses the risk of the wrong classification correctly into a target function and the parameters are adjusted and optimized based on this target function. The result of the proposed method is compared to SVR and decision tree in this section. Errors are calculated using the bellow equation. The proposed method excels by 99% to SVR and 81% to the decision tree method (Table 5).

$$Error_{methods} = \frac{1}{N} \sum_{i=1}^{n} e_i^2 \qquad (15)$$

*2) Second Experiment*

The map phase consists of two parts and a reduce phase has three parts. In this section, we compare the estimated weights in these two stages by the presented method, LATE and ESAMR. For this section, the stored executive information including the amount of processed data and the extent to which tasks are progressed are saved in a file and provided to the neural network as training information. The value of k is 10 for the ESAMR algorithm while the proposed method uses backpropagation to find weights. Results show 85% improvement compared to ESAMR and 99% to the LATE. Table 6 contains all the weight estimation of the proposed method compared to ESAMR and LATE algorithms. There are two numbers in each cell of this table. One represents the actual weight and the other one is the estimated one by the mentioned algorithm. It is essential to have these two numbers as close as possible in order to increase the accuracy of the method by correctly detecting straggler tasks.

*3) Third Experiment*

The purpose of the speculative execution is reducing the execution time. As a result, this experiment considers the remaining time estimation in the proposed method. First, the obtained results from executing the word count program with different input values have been stored. Then 20 tasks (map and reduce) are selected to evaluate the estimation on execution time. Experiments of applying ESAMR[36] are executed on six machines in which one of them is the manager and others are working nodes. Experiments of this research are performed on a Hadoop cluster consisting of five machines (one manager and four working nodes). Figure 4 and figure 5 show the difference between the estimated time and actual execution for Map and Reduce phases. As it is known, the proposed method is the closest case and has the closest estimate to the actual runtime and the lowest runtime among other methods. Based on Figure 6, the average declination in the error rate in the proposed method is 55% compared with the ESAMR and 77% improvement compared to LATE. Correct estimation of tasks remaining time is the key advantage of our method compared to the existing methods. The average distance between the real and estimated weights for the Map phase of LATE, ESAMR and our method is 68.3 33.85 18.1 and for Reduce phase is 176.75 89.95 27.3 respectively. ESAMR has better accuracy compared to LATE and to show how good the proposed method is to compare to ESAMR in Word Count, we have added figure B in Appendix which a calculation of difference estimation error for each task in Map and Reduce estimated by ESAMR and our method. Points above the line y=0 represent the better performance of our method compare to ESAMR.

`

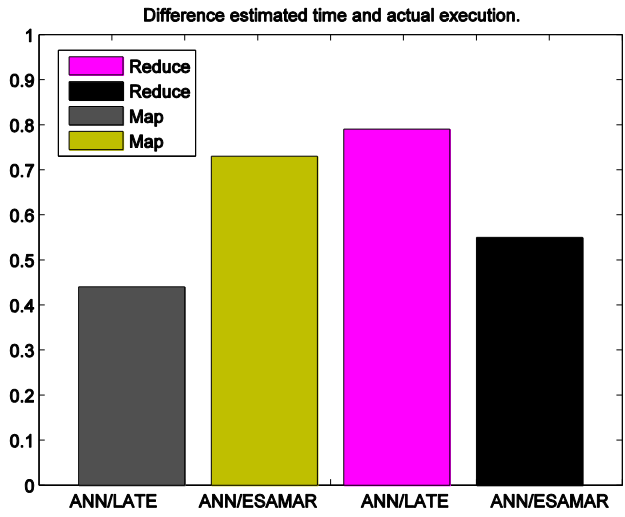

FIGURE 4: IMPROVEMENT IN ERROR HANDLING COMPARE TO PREVIOUS STUDIES.

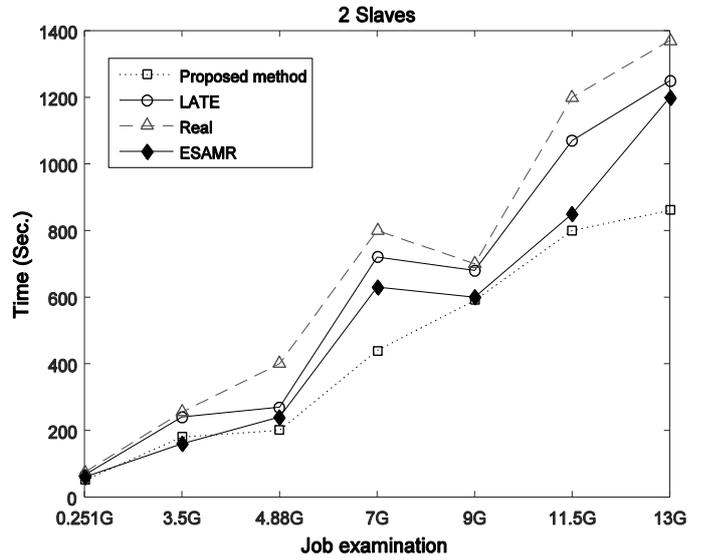

FIGURE 5: RUNTIME WITH TWO SLAVES.

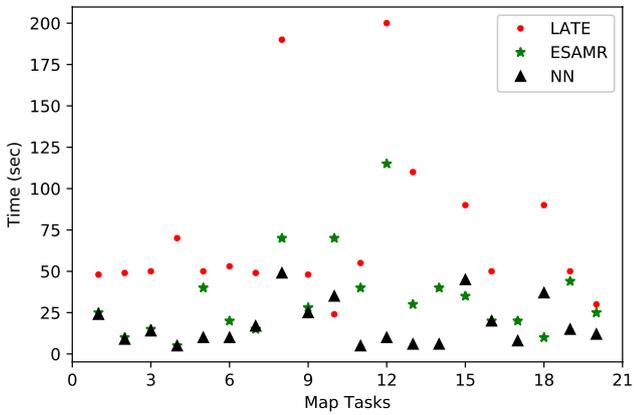

FIGURE 6: DIFFERENCE IN THE ESTIMATED RUNTIME IN MAPPING PHASE(WORDCOUNT).

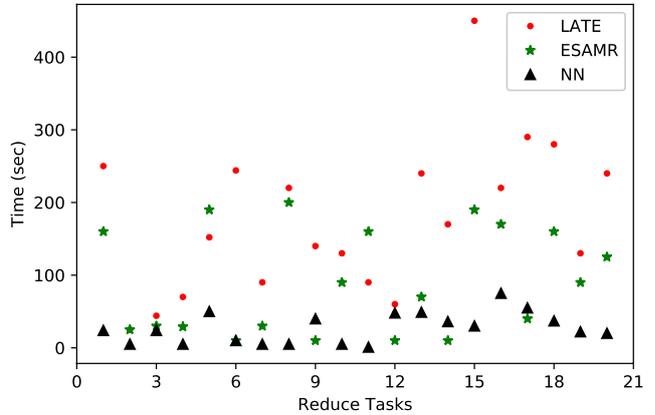

FIGURE 7: DIFFERENCE IN ESTIMATED RUN TIME IN REDUCE PHASE(WORDCOUNT).

TABLE 5.  WEIGHT ESTIMATION USING DIFFERENT ALGORITHMS.

A: LATE Algorithm

| Real amount | Estimated value |
|---|---|
| 0.78 | 0.77 |
| 0.214 | 0.181 |
| 0.28 | 0.211 |
| 0.147 | 0.145 |
| 0.93 | 0.94 |

B: SVR Algorithm

| Real amount | Estimated value |
|---|---|
| 0.46 | 0.91 |
| 0.66 | 0.893 |
| 0.34 | 0.592 |
| 0.25 | 0.231 |
| 0.16 | 0 |

C: Neural Network Algorithm

| Real amount | Estimated value |
|---|---|
| 0.76 | 0.811 |
| 0.449 | 0.462 |
| 0.616 | 0.611 |
| 0.235 | 0.213 |
| 0.561 | 0.512 |

`

TABLE 6.    WEIGHT ESTIMATION USING DIFFERENT METHODS VS REAL WEIGHTS OF A WORDCOUNT FOR SECOND EXPERIMENT.

A: Proposed method

| R1 | R2 |
|---|---|
| 0.78571-0.70432 | 0-0.0001 |
| 0.7-0.65 | 0-0.002 |
| 0.9019-0.8991 | 0.09803-0.0964 |
| 0.896551-0.8855 | 0.00086-0.0001 |
| 0.8544-0.8539 | 0.08181-0.0801 |
| 0.90277-0.9035 | 0.0532-0.058 |

B: ESAMR method

| R1 | R2 |
|---|---|
| 0.8538-0.7425 | 0-0.021 |
| 0.46-0.587 | 0-0.0002 |
| 0.8604-0.7401 | 0.0346-0.0158 |
| 0.9941-0.8954 | 0.0054-0.0023 |
| 0.7541-0.7451 | 0.045-0.0328 |
| 0.9921-0.8971 | 0.0641-0.0494 |

C: LATE algorithm

| R1 | R2 |
|---|---|
| 0.33-0.7432 | 0.33-0.08062 |
| 0.33-0.0001 | 0.33-0.04072 |
| 0.33-0.65 | 0.33-0.0792 |
| 0.33-0.002 | 0.33-0.06327 |
| 0.33-0.8991 | 0.33-0.05028 |
| 0.33-0.8991 | 0.33-0.05028 |

TABLE 7.    SUMMARY OF IMPROVEMENT IN PROPOSED METHOD.

| A: Number of experiment , B: Explanations , C: Percentage improvement compared to ESAMR , D: Percentage improvement compared to LATE , E: Percentage improvement compared to other methods | | | | |
|---|---|---|---|---|
| A | B | C | D | E |
| 1 | In this experiment, estimation of methods other than basic methods are compared | | | Compared to SVR 99% Compared to decision tree 81% |
| 2 | In this experiment, estimated weights of neural networks are compared with the estimated weights of the two basic methods | 85% | 99% | - |
| 3 | In this experiment, the estimated error rate of tasks' execution time compared with two basic methods are calculated | 55% | 77% | - |
| 4 | In this experiment, the execution time of job with the proposed algorithm is compared with the execution time of two basic methods | 15% | 24% | Compared to REAL 35% |

*4) Fourth Experiment*

This experiment considers the effects of data and the number of slave nodes on execution time. For this purpose, the word count program has executed with different inputs in the same conditions with all three algorithms. The obtained results with two slave nodes are depicted in Figure 7, the results with three slave nodes in Figure 8 and the results with four slave nodes are shown in Figure 9. As demonstrated, increasing the number of nodes will not always decrease execution time. It is due to the increasing data transmission time between nodes, and therefore, as the figure 10 also shows the increase of nodes in high volumes is cost-effective and in low-volume data, it does not increase or change the execution time. The remaining time of execution in the proposed method is estimated at 35% compared to real-time and 15 times compared to ESAMR. By timely identification of straggler tasks and assigning them to another node, the execution time decreases. Table 7 summarizes improvements in performed experiments.

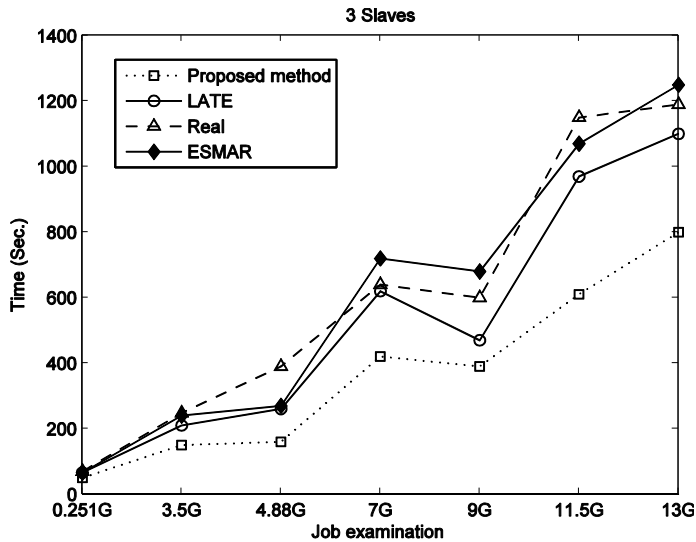

FIGURE 8: RUNTIME WITH THREE SLAVES.

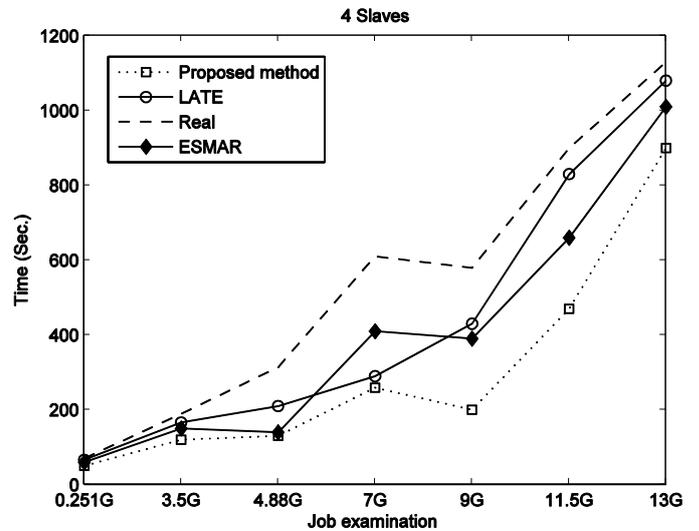

FIGURE 9: RUNTIME WITH FOUR SLAVES.

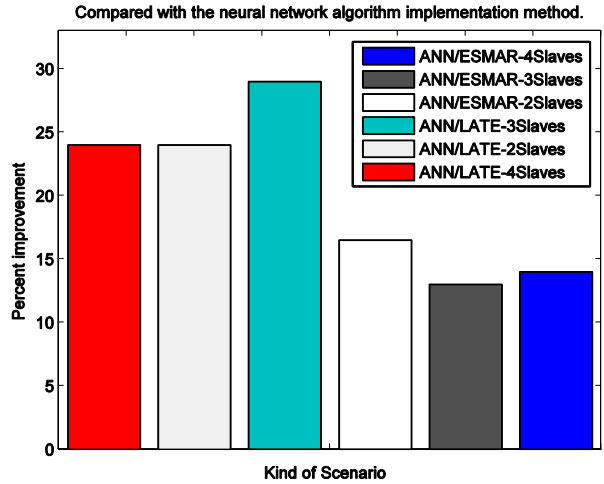

FIGURE 10: PERCENT IMPROVEMENT COMPARED WITH THE NEURAL NETWORK ALGORITHM IMPLEMENTATION METHOD AND LATE ESAMR.

*5) Qualifying the proposed method using Sort Benchmark*

Figures 11 and 12 are a comparison demonstration of the time to end estimation error of Map and Reduce tasks by ESAMR, LATE and the proposed method on a Sort 10GB job. Although the ESAMR has a small error on different tasks, the proposed method is doing better in the majority of tasks. With ESAMR, the differences between estimated and actual time to end of Map and Reduce tasks are 2 and 8 seconds respectively while the error in our method is 2 and 7 seconds respectively. The average distance between the real and estimated weights for Map phase of LATE, ESAMR and the proposed method is 17.9, 3.25, 2.95 and for Reduce phase is 126.55, 9.4, 9.25 respectively. As ESAMR has similar results to the presented model, we have added figure C in Appendix to demonstrate a comparison of estimation error for the proposed method and ESAMR for Sort benchmark. That is, we calculated the difference between the estimation error for each task in Map and Reduce estimated by ESAMR and the proposed method. Points above the line y=0 represent the better performance of the proposed method compare to ESAMR.

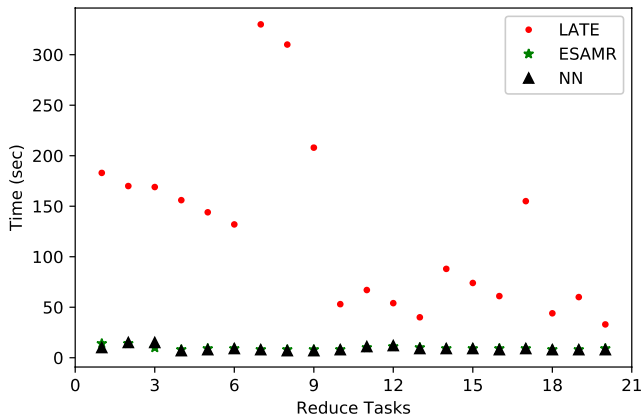

FIGURE 11: DIFFERENCE IN THE ESTIMATED RUNTIME IN MAPPING PHASE(SORT).

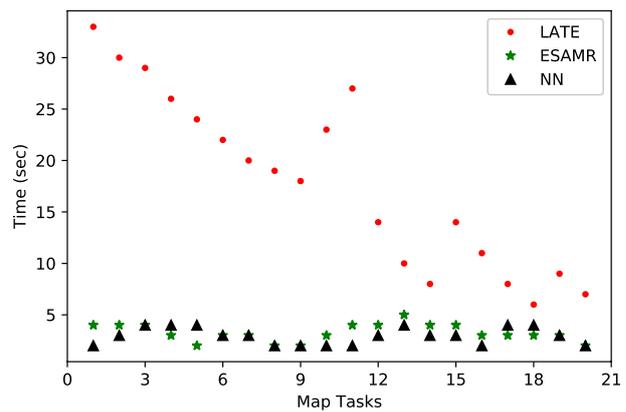

FIGURE 12: DIFFERENCE IN ESTIMATED RUN TIME IN REDUCE PHASE(SORT).

## V. CONCLUSIONS AND FUTURE DIRECTIONS

In Big Data processing, the speed of data analysis is of great significance. The purpose of this study is to improve the efficiency of the computational infrastructure of Big Data to accelerate data processing by speculative execution to identify straggler tasks. To achieve this goal, Speculative Execution with the ANN algorithm is proposed. With this development, the weight estimation improved by 86% compared to ESAMR and 99% compared to the LATE. Also, execution time using the proposed method improved by 24% and 15% compared to LATE and ESAMR, respectively.

Regarding the importance of short execution time, the proposed method can be further developed in the future. For example, to improve the process of speculative execution, the neural network can be used to identify the appropriate node for running a straggler task. For this purpose, other information such as the number of failures in different phases can also be used to determine the number of input data. In the proposed method, the support task starts from the start, and to process faster, the execution can be done by continuing the execution of the straggler task. It is better to consider the number of previous failures of a node before assigning a task. A combination of heuristic and artificial intelligence algorithms can be used in future work to process the data in parallel and as a result, to execute tasks faster. In this article by using NN, we estimate the remaining time of tasks, recognize the straggler tasks and estimation of weights. It is also a good idea to consider the following items in the estimation phase: 1) finding straggler tasks and the busy node which causes a delay in the system; 2) computing the remaining time of tasks in a proactive way. In this way, the system stores the behavior of tasks base on their characteristics and uses them in the case of new incoming tasks to estimate the remaining time more accurately; 3) using the Markov prediction model in estimating execution time and straggling tasks. This Markov model considers Map and Reduce phases in the past, the amount of processed data, and the number of executed tasks in a time period, estimating each stage weight, errors in estimating each stage weight and the node's remaining time in executing tasks. For future works, it is recommended to use other estimation methods as Taylor, learning automata and reinforcement learning.

### ACKNOWLEDGMENT

This work is supported in part by the National Natural Science Foundation of China under Grants 61632009 & 61472451, in part by the Guangdong Provincial Natural Science Foundation under Grant 2017A030308006 and Hgh-Level Talents Program of Higher Education in Guangdong Province under Grant 2016ZJ01.

`

`

## VI. APPENDIX

1. If reduce has not started
2. For each running task:
3. M1=CalculateMapTasksProgressScore
4. Else
5. <R1,R2>=CalculateReduceTasksWeightWithANN()
6. R3=1-R1-R2
7. CalculateReduceTasksProgressScore
8. Estimated End Time

ALGORITHM A: CALCULATE PROGRESS SCORE OR WEIGHT FOR EACH TASK BASED ON CURRENT PHASE OF EACH TASK.

1. Nf=processed key/value pairs
2. Na=all key/value pairs
3. subPS=Nf/Na
4. TTE= CalculateReduceTasksWeightWithANN()
5. Return TTE

ALGORITHM B: CALCULATE PROGRESS SCORE FOR TASKS IN MAP PHASE.

1. Nf=processed key/value pairs
2. Na=all ley/value pairs
3. sumPS=Nf/Na
4. If R2 has not started
5. PS=R1*subPS
6. Else if R3 has not started
7. PS=R1+R2*subPS
8. Else
9. PS=R1+R2+R3*subPS
10. End if
11. End if
12. Return PS

ALGORITHM C: COMPUTE PROGRESS SCORE FOR TASKS IN REDUCTION PHASE.

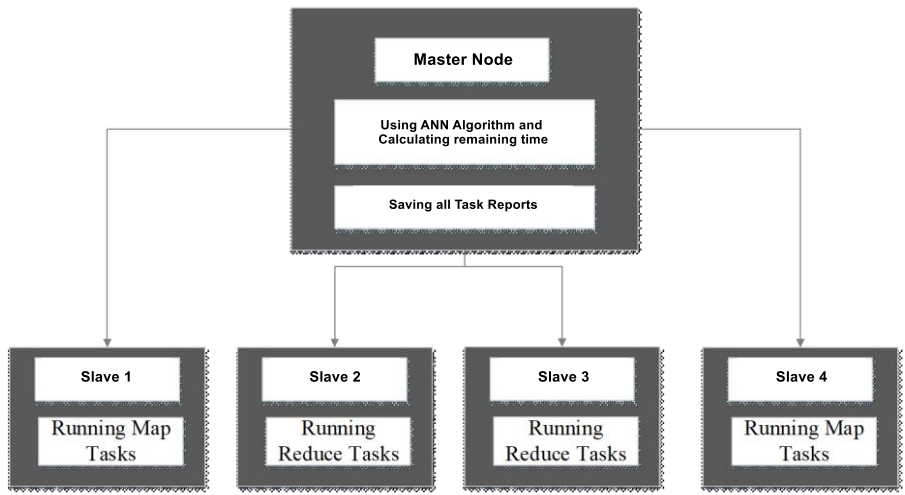

FIGURE A: RELATIONSHIPS OF NODES IN THE EXECUTIVE ENVIRONMENT

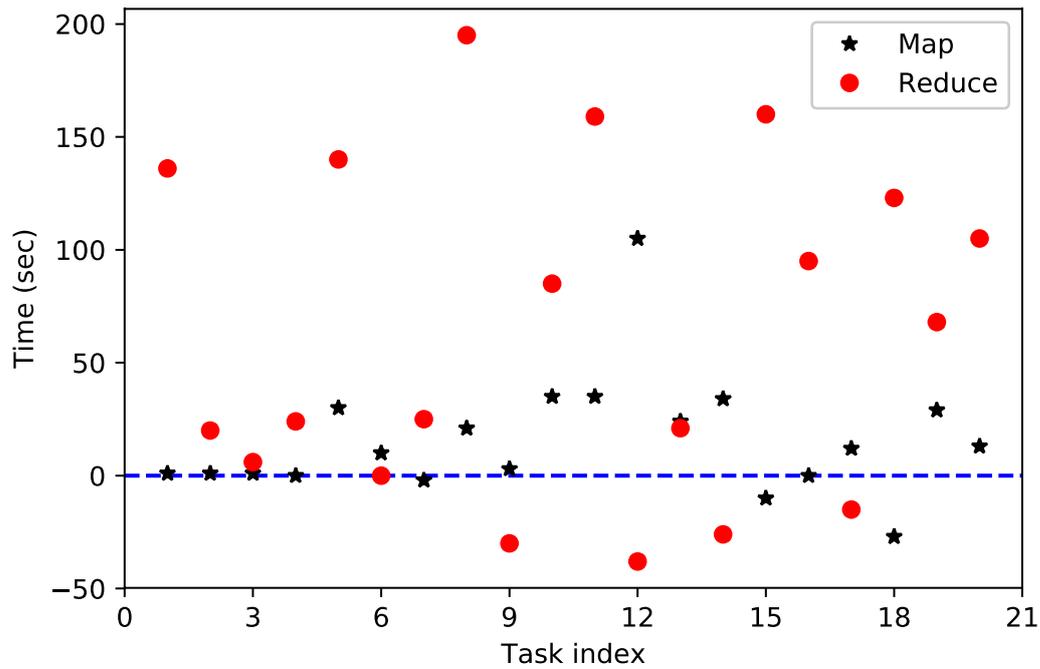

FIGURE B: DIFFERENCE OF ESTIMATION ERROR FOR ESAMR AND NN IN MAP AND REDUCE TASKS FOR NUMBERCOUNT BENCHMARK. POINTS HIGHR THAN THE DASHED LINE REPRESENT THE HIGHT ACCURACY OF OUR METHOD COMPARE TO ESAMR.

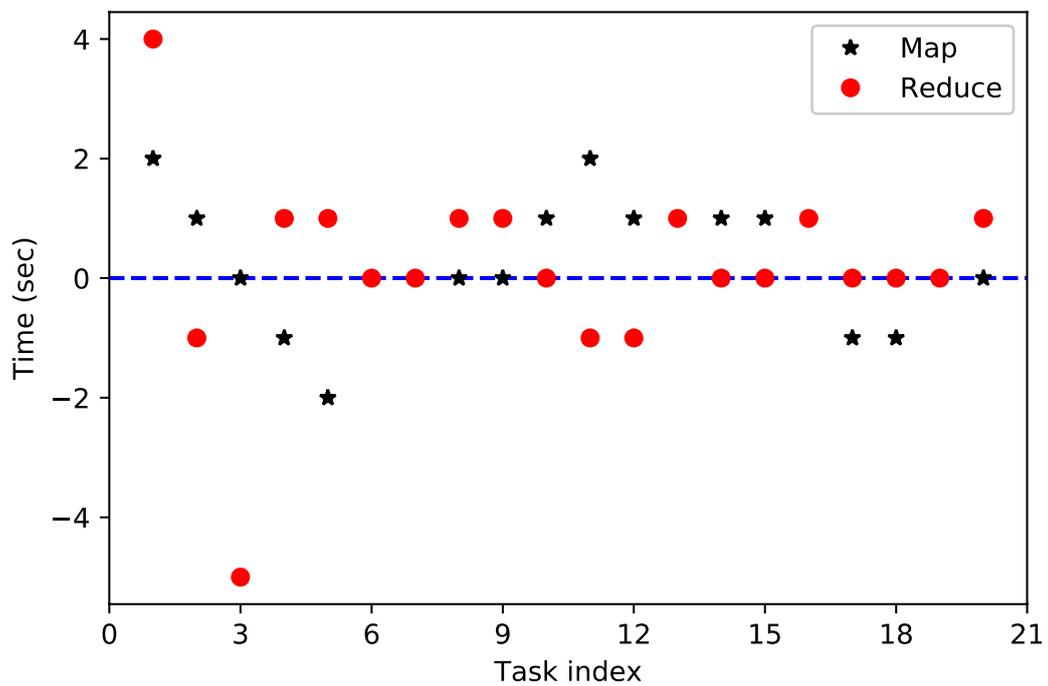

FIGURE C: DIFFERENCE OF ESTIMATION ERROR FOR ESAMR AND NN IN MAP AND REDUCE TASKS FOR SORT BENCHMARK. POINTS HIGHR THAN THE DASHED LINE REPRESENT THE HIGHT ACCURACY OF OUR METHOD COMPARE TO ESAMR.